\documentclass[12pt,preprint]{aastex}

\usepackage{amsmath}

\begin{document}

\title{On the role of supernova kicks in the formation of Galactic double neutron star systems
}
\author{
Yong Shao$^{1,2}$ and Xiang-Dong Li$^{1,2}$}

\affil{$^{1}$Department of Astronomy, Nanjing University, Nanjing 210046, China; shaoyong@nju.edu.cn}

\affil{$^{2}$Key Laboratory of Modern Astronomy and Astrophysics (Nanjing University), Ministry of
Education, Nanjing 210046, China; lixd@nju.edu.cn}

\begin{abstract}

In this work we focus on a group of Galactic double neutron star (DNS) systems with long orbital periods of 
$ \gtrsim 1$ day and low eccentricities of $\lesssim  0.4$. The feature of these orbital parameters is used to 
constrain the evolutionary 
processes of progenitor binaries and the supernova (SN) kicks of the second born NSs. Adopting that the mass
transfer during primordial binary evolution is highly non-conservative (rotation-dependent), the formation 
of DNS systems involves a double helium star binary phase, the common envelope (CE) evolution initiates before
the first NS formation. During the CE evolution the binary orbital energy is obviously larger when using a helium
star rather than a NS to expel the donor envelope, this can help explain the formation of DNS systems with long periods. 
SN kicks at NS birth can lead to eccentric orbits and even the disruption of binary systems, the low 
eccentricities require that the DNSs receive a small natal kick at the second collapse. 
Compared with the overall distribution of orbital parameters for observed DNS binaries, we propose that the second born NSs in
most DNS systems are subject to small natal kicks with the Maxwellian dispersion velocity 
of less than  $ 80 \,\rm km\,s^{-1} $, which can provide some constraints on the SN explosion processes.   
The mass distribution of DNS binaries is also briefly discussed.
We suggest that the rotation-dependent mass transfer mode and our results about SN kicks should be applied to 
massive binary evolution and population synthesis studies.

\end{abstract}


\section{Introduction}

More than 40 years has passed since the discovery of the first double neutron star (DNS) system 
PSR B1913+16 \citep{ht75}, and there are now over a dozen known in our Galaxy (see Table 1). The
orbital periods $ P_{\rm orb} $ of the DNS binaries are distributed in a wide range of $ \sim 0.1- 45 $ days, 
around half are close binaries ($ P_{\rm orb} < 0.5$ day) and the other half wide 
binaries ($ P_{\rm orb} > 1$ day). The orbital eccentricities $ e $ vary in a range
of $ \sim 0.1-0.8 $, and $ \sim 80\% $ of them have a relatively low value of $ e < 0.4 $. 
The components of the DNS systems  generally have a similar mass of  $ \sim 1.2-1.4 M_{\odot} $ 
except for PSR J0453+1559.
The observed pulsars are usually mildly recycled, with the spin periods $ P_{\rm spin} $ 
in the range of $ 17-185 \rm\, ms$ and the surface magnetic fields of the order
$ 10^{8}-10^{10} \rm\,G$.
 
Many previous works \citep[e.g.,][]{fv75,ty93,l97,dp03,i03,t17} have investigated the formation and evolution
of DNSs which involve a variety of binary evolutionary stages. The canonical channel for the DNS formation 
has already been established \citep[][for reviews]{bv91,tv06}. In a primordial  binary with both components
more massive than $ 8 M_{\odot} $, the primary star evolves to 
firstly fill its Roche lobe and transfer material to the secondary star. After that the primary becomes a helium star,
which finally explodes with a supernova (SN) to be a NS. Then the secondary star evolves and transfers mass to the NS, and the
binary becomes a high-mass X-ray binary (HMXB). Because of the large mass ratio, the mass transfer is dynamically unstable 
and binary system enters a common envelope (CE) phase \citep[see recent review by][]{i13} . During the CE phase, the orbital energy 
of the NS is used to dispel the donor's envelope, and the orbital period is dramatically decreased. As the stripped donor (a helium star) 
finally becomes a NS through SN explosion, a DNS is formed if the binary is not disrupted. 
The subsequent evolution of the DNS system is driven by gravitational wave radiation, and its orbital decay can  
lead to a DNS merger event associated with a short gamma-ray burst \citep{a17}. 

Modelling the DNS formation is subject to a number of uncertainties, especially those in the processes
of CE evolution, SN explosions and related natal kicks. CE evolution plays a vital role in the orbital evolution 
\citep{w84}. During SN explosions the natal kicks imparted to newborn NS can cause the orbit to be eccentric or
even disrupt the binary. It has been shown that most of the massive binaries are disrupted 
if the first NS receives a sufficiently large kick, and very few systems can survive to become HMXBs \citep{bz09,sl14}. 
From the orbital parameter distribution of the observed HMXBs, \citet{p02} suggested that the NSs
in wide and nearly circular binaries likely received small natal kicks, implying that they
 originated from electron-capture supernovae \citep[ECS,][]{p04,v04} rather than core-collapse supernovae (CCS).
After the second SN explosion, the survived DNSs are generally in eccentric orbits. Based on the distributions
of the DNS masses and their orbital parameters, \citet{bp16} proposed that the second collapse in the 
majority of the observed DNS systems involves small mass ejection and a low kick velocity \citep[see also][]{t17}. 

Since the recent discoveries of gravitational wave transients by the advanced LIGO detector \citep[e.g.][]{a16,a17}, 
a study of gravitational wave astronomy enters a golden age. Based on massive binary evolution, the 
rates of the merger events in the local universe can be theoretically estimated by a binary population synthesis method.
However, the input parameters and physical assumptions should be carefully verified before the method is used to predict
the merger rates in the local universe. Population synthesis studies 
\citep[see e.g.,][]{py98,a15,db15,b18,c18,k18,v18,gm18a,gm18b,mg18} can be used to evaluate these uncertainties 
by adopting different input parameters to match the calculated distributions of the DNS population with those in our Galaxy. 
In this paper, we model the population of the Galactic DNSs. Compared to previous works, 
we have updated some important treatments of binary evolution, paying particular attention to 
the stability of mass transfer in primordial binaries, 
CE evolution, SN explosions and natal kicks. Our results may provide helpful constraints on the input 
physics of massive binary evolution.
The remainder of this paper is organized as follows. In Section 2, we introduce the binary population synthesis
method and the adopted assumptions. We present our calculated results and discussions in Section~3. 
We make a comparison with previous works in Section~4 and give a summary in Section~5.

\section{Binary population synthesis method}

To explore the distribution of the Galactic DNS systems, we use the population synthesis code \textit{BSE}
originally developed by \citet{h02}. With \textit{BSE} we can simulate the evolution of a large of binary
stars with different initial parameters. The modelling of binary evolution involves detailed treatments of
stellar evolution, tidal interactions, mass and angular momentum loss, and mass and angular momentum transfer. 
In the code some modifications have been made
by \citet{sl14}. Several important physical inputs are described as follows.

\subsection{Mass transfer in the primordial binaries}

Mass transfer begins in a primordial binary when the primary star evolves to fill its Roche lobe.
The mass transfer efficiency (i.e. the fraction of the accreted matter by the secondary among the 
transferred matter) is a vital factor to determine the fate of the binary system.  Rapid mass
accretion can drive the secondary star to get out of thermal equilibrium and the responding expansion 
may also cause the secondary star to fill its own Roche lobe, leading to the formation of a contact 
binary \citep{ne01}. The contact binary will evolve to merge if the primary is a main sequence
star, otherwise the binary system will enter a CE phase \citep[see e.g.,][]{d13,sl14}. 

\citet{p81} showed 
that only a small amount of accreted mass can accelerate a star to reach its critical rotation.
When the secondary star is a rapid rotator, the mass transfer efficiency is subject to large 
uncertainty \citep{l98,p05,se09}. When modelling the evolution of the primordial binaries,
\citet{sl14} built three mass transfer modes. In Mode I, the mass accretion rate onto the secondary star is 
assumed to be dependent on its rotating rate, equal to the mass transfer rate multiplied by a factor of
$ 1-\Omega/\Omega_{\rm cr}$ (where $ \Omega $ and $ \Omega_{\rm cr} $ are the angular velocity of the 
secondary and the corresponding critical value, respectively); in Mode II, the mass accretion efficiency is 
simply assumed to be 0.5; in Mode III, the mass accretion rate is assumed to be the mass transfer rate multiplied by 
a factor of $ \min (10\frac{\tau_{\dot{M}}}{\tau_{KH2}},1)$, where $\tau_{\dot{M}}$ is the mass transfer 
timescale and $\tau_{KH2}$ the secondary$'$s thermal timescale \citep{h02}. 
In the case of rapid mass accretion, the secondary star will get out of thermal equilibrium, expand
and become overluminous. The thermal timescale of the secondary is greatly decreased, so
the mass transfer process in Mode III is generally quasi-conservative, which is disfavored in massive 
binary evolution \citep{sl14} and will not be considered in this work. In comparison the mass transfer 
efficiency  in Mode I can be $ \lesssim 0.2 $ for Case B binaries \citep{sl16}. The maximal initial mass ratios
for avoiding the contact phase can reach as high as $ \sim 6 $ in Mode I, and 
drops to $ \sim 2 $ in Mode II and III \citep{sl14}. 

In our calculation we adopt Mode I to treat the mass transfer in the primordial binaries,
unless specified otherwise. 
The reasons are briefly described as follows.
When modelling the evolution of massive binaries, \citet{sl16} demonstrated that the 
Galactic Wolf-Rayet/O binaries can be produced only if the mass transfer 
is highly non-conservative. It is also noted
\citet{d07} discussed the mass transfer efficiency in massive close binaries and found a tendency
that systems with long initial orbital periods evolve in a  more non-conservative way than narrow systems,
which might be caused by the coupling of spin-up and tidal interaction during the mass transfer 
phase.

Highly non-conservative mass transfer in  Mode I can prevent the binary entering a CE phase and result in 
the formation of a long period system \citep[see Figure 6 of][]{sl14}. 
MWC 656 is the first confirmed Be/black hole binary with an orbital period of $ \sim 60 $
days \citep{c14}.  Assuming that  the rapid rotating Be star has been accelerated by stable mass transfer
during the primordial binary evolution, \citet{sl14} showed that non-conservative mass transfer is more likely 
to lead to the formation of MWC 656-like binaries. 
Recently \citet{k18} suggested that the binary AS 386 may be a candidate Be/black hole system, which
contains a Be star of mass $ 6-8 M_{\odot} $ in a $ \sim 131 $ day orbit. 
Such a long period system is hardly produced if the progenitor binary has experienced CE evolution \citep{p03,g15}.

By comparing the fraction of runaway Be stars among the whole Be star population, \citet{be18} concluded that 
Be stars could be explained by an origin of mass transfer in binaries, that is, most of the runway Be stars 
are the products of disrupted Be/NS binaries. For the identified
runway Be stars \citep{bg01}, the stellar spectral types can be late to be B8 (the corresponding stellar mass 
$ \sim 3M_{\odot} $). This feature can be well produced by inefficient mass accretion which has spun up 
the secondary star without significant mass growth.

Now that we assume that the mass transfer is highly inefficient, 
for the matter that is not accreted by the secondary star, we let it escape the binary system 
in the form of isotropic wind, taking away the specific angular momentum of the secondary star. In addition,
we follow \citet{h02} to treat the process of rejuvenation, during which the secondary star appears even younger
when fresh hydrogen is mixed into its growing convective core due to mass accretion. 

\subsection{Supernova explosion mechanisms and natal kicks}

A NS can be formed through both ECS and CCS. 
In the \textit{BSE} code, the helium core mass at the base of 
the asymptotic giant branch is used to set the limits for identifying various CO cores \citep{h02}. 
If the helium
core mass is lower than $ 1.83M_{\odot} $, the star will form a degenerate CO core and finally
leave a CO white dwarf. Above this mass the star develops a partially degenerate CO core.
If the core reaches a critical mass of $ 1.08M_{\odot} $, it will non-explosively burn into a 
degenerate ONe core. Stars with an ONe core more massive than $ 1.38M_{\odot} $ are believed to collapse 
into NSs through ECS, otherwise form an ONe white dwarf.  
When the helium core is massive 
enough to form a non-degenerate CO core, stable nuclear burning will continue until a CCS occurs. 
However, investigations show that the trigger of ECS explosions is subject to big uncertainties, 
and mass transfer in binary systems makes it more complicated \citep[see e.g.,][]{n84,p04,wh15,p17}. 
In the calculations, we consider several possible helium core mass ranges $ \Delta M_{\rm ecs} $
($ 1.83-2.25 M_{\odot} $, $ 1.83-2.5 M_{\odot} $, $ 1.83-2.75 M_{\odot} $),
 within which the star may eventually explode in 
an ECS explosion. The upper masses of these ranges can directly distinguish whether
a star ends its life in an ECS or a CCS explosion. The larger the mass range, the more NSs formed 
through ECS.

We adopt the rapid SN model of 
\citet{f12} to deal with the mass of NSs formed from CCS, which varies in the range of $ \sim 1.1 - 2M_{\odot} $.
The gravitational mass of ECS NSs is set to be $ 1.3M_{\odot} $ considering the fact that some baryonic mass of 
the $1.38  M_{\odot} $ ONe core is converted into the binding energy of the NS \citep{t96,f12}.
A newborn NS is assumed to be imparted by a natal kick, which may lead to the disruption of the binary system.
Based on the observations of the pulsars' proper motions, \citet{h05} demonstrated that the pulsar birth velocities
can be described by a Maxwellian distribution with a dispersion velocity of $\sigma = 265 \rm\, km\, s^{-1} $.
More recently, \citet{vic17} argued that the birth velocities of pulsars can be better fitted by a bimodal distribution
with $\sigma \simeq 80 $ and $ 320 \rm\, km\, s^{-1} $, which may respectively be related to
ECS and CCS. Analysing the nature of observed DNSs reveals that the second formed NSs
in most systems are likely to receive low kicks \citep{bp16,t17}. In our calculations, the NS kick velocities are assumed
to obey Maxwellian distributions. We adopt $\sigma_{\rm k,ecs} = 20 $, 40 and $80 \rm\, km\, s^{-1} $ for ECS NSs  
and $\sigma_{\rm k,ccs} = 150 $ and $300 \rm\, km\, s^{-1} $ for CCS NSs to examine their 
possible influence on the DNS formation. 

\subsection{Mass transfer in NS binaries}

After the birth of the first NS, the binary evolves into the HMXB phase.
Modelling the evolution of NS HMXBs reveals that the maximal mass ratio for stable mass transfer
is $ \sim 3.5 $ \citep{t00,prp02,sl12}. If the donor is a helium star, we
assume that the binary enters CE evolution only when the orbital period is less than 0.06 day \citep{t15}.
In the case of unstable mass transfer that leads to CE evolution, we calculate the orbital energy of the binary system 
to examine whether it is sufficient to expel the stellar envelope.
We use the results of \citet{xl10} and \citet{w16} for the binding energy parameter $ \lambda $ of the envelope, 
and take the CE efficiency $ \alpha_{\rm CE} $ to be 1.0 in the calculations. 

For stellar wind mass loss, we employ the fitting formula of \citet{nd90}, except for hot OB stars,
for which we use the simulated rates of \citet{v01}. For the stripped helium stars and Wolf-Rayet stars, we adopt 
the rates of \citet{v17}\footnote{ Note that the wind mass loss rates of \citet{v17} are recommended only for stripped
stars through the interaction with a companion, and may not be suitable for classical WR stars.}.  
During the CE phase, we ignore any mass accretion because of its very short duration. In NS binaries
with a helium donor star, mass transfer proceeds rapidly for a short ($ < 10^{5}  \,\rm yr$) time
through Roche lobe overflow \citep{t15}, the accretion rate of the NS is assumed to be limited by its Eddington rate.

\section{Results and discussions}

We have performed a series of Monte Carlo simulations for a large number of
binary systems with different initial parameters. For each of the considered models, we simulate the evolution
of $ 10^{8} $ binary systems. We follow \citet{h02} to set the initial parameters
of the primordial binaries. The primary masses obey the initial mass function given by \citet{k93}, and the
mass ratios of the secondary to the primary satisfy a flat distribution between 0 and 1. The distribution of 
the orbital separations is assumed to be flat in the logarithm in the range of $ 3-10^{4} R_{\odot}$. 
All binaries are assumed to have circular orbits,
since the outcome of the interactions of binary systems with the same semilatus rectum is almost independent 
of eccentricity \citep{h02}. Several authors \citep{s78,d06,rw10} have proposed that the star
formation rate in our Galaxy is $ \sim1-5 M_{\odot}\,\rm yr^{-1} $. Here we adopt a moderate rate of 
$ 3 M_{\odot}\,\rm yr^{-1} $ over a period of 10 Gyr. When investigating the star formation history
of the Galaxy, \citet{r00} showed that the star formation rate is not likely to
change by much more than a factor of two over the past 10 Gyr. So adopting a constant star formation rate 
should be an acceptable approximation.

From each of the population synthesis calculations, we obtain a few to several ten thousand initial DNS systems.
The corresponding  birthrates can be calculated according to the parameters of their 
primordial binaries \citep{h02}. After the DNS formation, we adopt the formulae of \citet{p64} to trace
the subsequent orbital evolution with time due to gravitational wave radiation.
The DNS merger rate  within the lifetime of the Galaxy is mainly dependent on the initial orbital 
parameter of the DNS binary\footnote{The DNS formation timescale is $ \sim 10^{7} $ yr and negligible compared with the 
timespan of 10 Gyr.}. In the calculations, we assume that the DNS binaries evolve to merge when the orbital periods 
shrink to be less than 0.001 day. Since we adopt a constant star formation rate of $ 3 M_{\odot}\,\rm yr^{-1} $, 
the birthrate of the Galactic DNS systems does not vary with time, 
while the merger rate tends to gradually increase  over the past 10 Gyr.
The total number of the DNS binaries in our Galaxy
can be obtained by accumulating the product of the birthrate and the evolutionary time for every binary. 
In Table 2, we show the merger rate $ R_{\rm merger} $, the birthrate $ R_{\rm birth} $ and   
the total number of the Galactic DNSs at present, and the fraction $ f $ of DNS systems with 
the second NS being formed through ECS among all DNSs. Obviously the larger values of 
both $ \sigma_{\rm k,ecs} $ and $ \sigma_{\rm k,ccs} $ result in the decreases of
$ R_{\rm merger} $ and $ R_{\rm birth} $. We obtain $ R_{\rm merger} \sim 2-10 \,\rm Myr^{-1} $,
 $ R_{\rm birth} \sim 3-27 \,\rm Myr^{-1} $, and the total number of DNSs in 
our Galaxy is of the order  $ 10^{4} $. It is also seen that wider $ \Delta M_{\rm ecs} $ leads to higher $ f $, 
which varies in the range of $ 0.06-0.85 $.

\subsection{The $ P_{\rm orb}-e $ diagram}

Since the orbital evolution of DNS systems is solely driven by gravitational wave radiation, 
we can synthesize the orbital parameters of the simulated DNS population into a diagram that 
represents the snapshot after 10 Gyr star formation and evolution.
We show in Figure~1 the calculated distribution of Galactic DNS systems (with the numbers 
given in Table 2) in the $ P_{\rm orb}-e $ plane when adopting 
$ \Delta M_{\rm ecs} = (1.83-2.25 M_{\odot}) $.
The left, middle and right panels correspond to  $\sigma_{\rm k,ecs} =$
20, 40 and $80 \rm\, km\, s^{-1} $ for ECS NSs, and the top and bottom panels correspond to $\sigma_{\rm k,ccs} =$  150 and 
$300 \rm\, km\, s^{-1} $ for CCS NSs,  respectively. Each panel includes $ 20\times 20 $ image elements,
in which $ P_{\rm orb} $ varies logarithmically from 0.01 to $ 10^{4} $ days with steps of 0.3, and $ e $ linearly from
0 to 1 with steps of 0.05. Until now there exist 14 observed DNSs whose orbital periods and eccentricities
are precisely measured (see Table 1). They are plotted as blue triangles in the figure. Figure 1 shows that the calculated DNSs are
mainly distributed 
in a region with relatively large eccentricities ($ e >0.4 $), so the observed systems with long orbital
periods ($ \gtrsim 1 $ day) and low eccentricities ($ \lesssim 0.4 $) can be hardly reproduced. The reason is that with
 $\Delta M_{\rm ecs} = (1.83-2.25 M_{\odot}) $  the second NSs in most systems 
are born in a CCS explosion with a high natal kick, which tends to increase the orbital eccentricities
and even disrupt the original systems (especially for wide binaries). 

Figures 2 and 3 show the similar distributions of Galactic DNSs in the $ P_{\rm orb}-e $ plane with 
$ \Delta M_{\rm ecs} = (1.83-2.5 M_{\odot} )$ and ($ 1.83-2.75 M_{\odot} $), respectively. In these two cases
the value of $ f $ is remarkably increased and  can reach as high as 0.85 
when $\sigma_{\rm k,ecs} = 20 \rm\, km\, s^{-1} $  and 
$\sigma_{\rm k,ccs} = 300 \rm\, km\, s^{-1} $ (see Table 2). The second born NSs in most systems are imparted to small 
natal kicks, and the long-period binaries can effectively avoid disruption. In addition, the predicted DNS binaries
tend to have low eccentricities. Generally the observed DNS binaries can be well covered by a high probability 
region in the $ P_{\rm orb}-e $ diagrams when taking $ \Delta M_{\rm ecs} = (1.83-2.75 M_{\odot} )$, 
expect the case with $\sigma_{\rm k,ecs} = 80 \rm\, km\, s^{-1} $ (see right panels). 
Consistent with previous works \citep[e.g.][]{bp16,t17}, our results suggest that the formation of the second 
NS be accompanied with a low natal kick. 

We then try to quantitatively compare the calculated results with observations to constrain 
the adopted models with different input parameters.
Although the observed Galactic DNS sample is 
limited by small size and the detection of such systems may be subject to several selection effects
\citep[see Section 2.1 of][]{t17},
we employ the Bayesian analysis used by \citet{a15} to  compare the goodness of match
for every model, which involves two independent orbital parameters of $ P_{\rm orb} $ and  $ e $. The
Bayes' theorem gives a probability expression of
\begin{equation}
 P(M\mid D) = \frac{P(D\mid M)P(M)}{P(D)},
\end{equation}
where $ P(M\mid D) $ is the posterior probability, i.e. the probability related to our model $ M$
given the observed data $  D$ (i.e., the data set of $ P_{\rm orb} $ and  $ e $),
$ P(D\mid M) $ is the likelihood of the observed data given that the model is true, which is denoted as $ \Lambda(D) $,
$ P(M) $ is the prior probability of our model, and
$ P(D) $ a normalizing constant that is independent of the model. As all models  have the same prior probability, it can be
absorbed into the normalizing constant as $ C =  P(M)/P(D)$.  So $P(M\mid D)$ can be rewritten as 
\begin{equation}
 P(M\mid D) = C \Lambda(D) .
\end{equation}
Since $ C $ is independent of the models, the value of $ \Lambda(D)  $ gives the relative probability of each model.
Because all of the observed 
DNSs in the sample are independent, the probability of the observed data given the model is equal to the product
of the probabilities of each binary system:
\begin{equation}
\Lambda(P_{\rm orb},e) = \prod_{i} P(P_{{\rm orb}, i},e_{i}\mid M).
\end{equation}
Note that the errors of the observed orbital period and eccentricity for each DNS system are very small and can be 
neglected. Then
$ P(P_{{\rm orb}, i},e_{i}\mid M) $ reflects the probability density of a model at a specific point in the 
$ P_{\rm orb}-e $ parameter space, which corresponds to a specific DNS. Adopting similar treatment of \citet{a15}, 
we calculate $ P(P_{{\rm orb}, i},e_{i}\mid M) $ by binning  the simulated systems into two dimensions with a bin size
of $ \Delta \log(P_{\rm orb}) = 0.3 $ in orbital period and $ \Delta e = 0.05  $ in eccentricity. Finally the 
derived value of $ \Lambda(P_{\rm orb},e)  $ is used to rank our models (see Table 2). 
The model that best matches the observations
involves the input parameters of $\sigma_{\rm k,ecs} = 40 \rm\, km\, s^{-1} $, $\sigma_{\rm k,ccs} = 300 
\rm\, km\, s^{-1} $ and $ \Delta M_{\rm ecs} = (1.83-2.75 M_{\odot}) $.

To display the effect of the mass transfer process during the primordial binary 
evolution on the DNS population, in Fig.~4 we present the corresponding distributions of DNSs in the 
$ P_{\rm orb}-e $ plane when the mass transfer efficiency is taken to be a constant value of 0.5 (i.e., Mode II).
The left, middle and right panels correspond to  $ \Delta M_{\rm ecs} = (1.83-2.25 M_{\odot}) $, 
$( 1.83-2.5 M_{\odot} )$ and $ (1.83-2.75 M_{\odot}) $, respectively. The kick velocity parameters are all set
to be $\sigma_{\rm k,ecs} = 40 \rm\, km\, s^{-1} $ and $\sigma_{\rm k,ccs} = 300 \rm\, km\, s^{-1} $.
Compared with the situations in Mode I,
the predicted distribution of the DNS systems shifts to a region with shorter orbital periods 
and seems not to match observations. The main reason is that the birthrate of close DNSs
in Mode II is greatly enhanced since a large number of original NS/white dwarf binaries
(post-CE products of HMXBs) in Mode I are transformed into DNS systems due to the
higher mass transfer  efficiency. Further, more efficient mass transfer can significantly increase
the secondary mass, so it takes more kinetic energy of the NS to expel the secondary's envelope 
during the CE evolution.

\subsection{The mass distribution of  DNS systems}

All of the observed Galactic DNSs except PSR J0453+1559 have similar masses of $ \sim 1.2-1.4 M_{\odot} $ (see Table 1).
As expected, recycled pulsars are generally more massive than their non-recycled companions
because a certain amount of mass is required to recycle a pulsar. In Figure~5 we present the distributions
of the pulsar mass $ M_{\rm p} $ (first formed NS, left), the companion mass $ M_{\rm c} $ (second formed NS, middle) and the 
total mass $ M_{\rm t} $ (right) 
for the Galactic DNS systems under different assumed models. The panels form top to bottom correspond to $\Delta M_{\rm ecs}
= (1.83-2.25 M_{\odot})$, $ (1.83-2.5 M_{\odot})$ and $(1.83-2.75 M_{\odot} )$, respectively.  
In each panel, the black, red, and blue curves correspond to 
$\sigma_{\rm k,ecs} = $  20, 40, and $80 \rm\, km\, s^{-1} $, and the dashed and solid curves correspond to 
$\sigma_{\rm k,ccs} = $ 150 and $300 \rm\, km\, s^{-1} $ respectively.  
The green curves represent the observed data multiplied by a factor of $ 10^{4} $. 
We find that the calculated mass distributions of both DNS components generally have two peaks of $ \sim 1.1 M_{\odot} $ 
\citep[minimum NS mass in the rapid SN model of][]{f12} and $ \sim 1.3 M_{\odot} $ (the mass of NSs formed from ECS),
in agreement with the results of \citet{gm18b} and \citet{mg18}.

The calculated mass distribution at the peaks of $ \sim 1.1 M_{\odot} $ and $ \sim 1.3 M_{\odot} $ 
could be used to distinguish the NSs formed from CCS and ECS, respectively. 
It is unsurprising that the percentage at each peak changes with $ \Delta M_{\rm ecs} $, the former decreases 
and the latter increases with increasing $ \Delta M_{\rm ecs} $ since more ECS NSs are produced. Even in the case of 
$\Delta M_{\rm ecs} = (1.83-2.25 M_{\odot})$, quite a fraction of DNS binaries undergo at least one ECS explosion.
This is consistent with the result of \citet{gm18a} for merging DNSs, indicating that the ECS is a fundamental process for the 
DNS formation. We can see that the overall distributions of calculated DNS masses, although in agreement with other studies, 
are actually not compatible with observations. The reason may be that the adopted SN mechanisms are still too simplistic, and 
there may not be one to one relation between the compact star mass and the core mass of their progenitors. 
Furthermore a NS can increase its mass due to the possible accretion during the recycling phase. 

In spite of the mass distribution for DNS binaries, comparison between the calculated and 
measured orbital parameters gives the best choice of the input
parameters of $\sigma_{\rm k,ecs} = 40 \rm\, km\, s^{-1} $, $\sigma_{\rm k,ccs} = 300 
\rm\, km\, s^{-1} $ and $ \Delta M_{\rm ecs} = (1.83-2.75 M_{\odot} )$ (see the rank of different models in Table 2), 
which are therefore set to be default input parameters in the following.

\subsection{Formation channels of DNS systems}

The formation channels of DNS systems have been explored by many authors.
Besides the canonical channel in which the binary experiences a HMXB stage \citep{bv91,tv06}, there is 
an alternative channel which involves a double helium star phase. In the double core CE scenario \citep{b95}
 a DNS system is the descendant of a double helium star binary, which is the result of 
a CE phase in a binary with both components of very similar masses. Both components of the binary have developed
a compact core at the beginning of mass transfer. Such a binary will enter the contact phase that is followed
by the CE evolution. The post-CE product is a double helium star binary in a relative close orbit. After 
experiencing two SN explosions, the survived binary is a DNS system \citep[see also][]{dps06}. 

By analysing the calculated results, we find that the double helium star binary can be produced in an another way. 
Here both of the binary components also initially have almost equal mass. The secondary star 
is still in the main sequence at the onset of mass transfer. The mass transfer proceeds stably between the
binary components, after which the primary leaves a helium star and the secondary approaches the end of
main sequence. The secondary then evolves to be a supergiant star and transfers mass to the helium star,
resulting in CE evolution.  
The outcome of CE evolution is a double helium star binary. A similar scenario was suggested by \citet{py98}, in which 
the mass exchange during the primordial binary evolution was assumed to be almost conservative, but the process of rejuvenation was 
not considered in their calculations. In the case of conservative mass transfer, this can cause the secondary to be younger, and  
 stay in the main sequence when the primary explodes to be a NS. 
A highly non-conservative mass transfer like in Mode I 
can effectively reduce the rejuvenation of the secondary. The primary helium star 
firstly explodes to be a NS, which is subsequently recycled by the mass transfer from its helium star companion.
When the secondary evolves to be a NS, the binary finally turns into a DNS system. 

Adopting the default model with the input parameters of  
$\sigma_{\rm k,ecs} = 40 \rm\, km\, s^{-1} $, $\sigma_{\rm k,ccs} = 300 
\rm\, km\, s^{-1} $ and $ \Delta M_{\rm ecs} = (1.83-2.75 M_{\odot}) $, we obtain the total birthrate of Galactic DNSs to be 
$ \sim 10\,\rm Myr^{-1} $ (see Table 2). The DNS binaries formed from the canonical channel
have a birthrate of $ \sim 4\,\rm Myr^{-1} $, and the two scenarios involving a double helium star 
binary contribute a similar birthrate of $ \sim 3\,\rm Myr^{-1} $. This means 
that the double helium star channel is likely to dominate the formation of Galactic DNS systems.

Figure~6 shows the calculated distribution of DNS systems in the $ P_{\rm orb}-e $ plane when  
taking into account different formation channels. 
The left and right panels correspond to the results from the canonical channel and the double helium star
channel, respectively. In the canonical channel, the DNS systems are dominated by binaries
with short orbital periods of $ \lesssim 1 $ day and low eccentricities of $ \lesssim 0.4 $, their distributions show similar 
features to that in Mode II (see the right panel of Figure~4). The double helium star
channel tends to produce the DNS systems with long periods of $ \gtrsim 1 $ day and high eccentricities  
$ \gtrsim  0.1 $. The difference originates from the fact that the binary orbital energy that is used to expel the CE in the former 
is considerably smaller than in the latter. Here the fraction of systems that formed from the canonical channel 
among all DNSs is $ \sim 0.3 $. 

In Figure~7 we present the calculated distributions of the DNS systems 
in the $ P_{\rm orb}-e $ plane, in which the second born NS originates from
ECS (left) or CCS (right) explosion. The NSs formed through CCS have received relatively large natal kicks, which 
tend to produce the DNS systems with large eccentricities. On the contrary, the DNS systems generally have low
eccentricities for ECS explosions. A large number of relatively wide systems can 
survive the SN explosions, and evolve to DNS systems with long orbital periods 
and low eccentricities.
The fraction of the DNS systems that experiences a second ECS explosion 
can be as high as 0.74 (see Table 2). We note that the PSRs J1757-1854 and B1913+16 have relative short periods and
large eccentricities, and their orbital decay timescales are of the order 100 Myr. These two binaries are 
not easily accounted for by the current scenarios.

\section{Comparison with other works}

In this section, we compare our results 
with recent population synthesis works of \citet{a15}, \citet{k18} and \citet{v18}.

The most important parameters in \citet{a15} include the combined CE parameter $ \alpha_{\rm CE} \lambda$,
the NS kick velocity distribution and the ECS mass range (same as our $\Delta M_{\rm ecs}$).
A significant difference between their work and ours is the treatment of the $ \alpha_{\rm CE} \lambda$, which
was set to be constant by \citet{a15}. Note also that the double helium star channel is not considered by \citet{a15}, who 
focus on the canonical channel.
To match the orbital parameters of the known eight DNSs, their models are required to produce wider binaries than 
our canonical channel. This indicates that their adopted $ \alpha_{\rm CE} \lambda$ is generally larger 
than ours. By testing different models to constrain the input parameters, they found that 
$ \alpha_{\rm CE} \lambda \lesssim 0.25$ is effectively ruled out, the allowed range is $ 0.3-1.0 $ and 
the best choice is 0.5. Considering the fact that the value of $ \lambda $ for massive supergiants are less than 0.1 
\citep{dt00,p03,w16}, this implies that the value of $ \alpha_{\rm CE} $ in \citet{a15} is significantly larger than unity.
After the discovery of PSR J1930-1852 with the longest known orbital period of 45 days, \citet{s15} argued that 
many models of \citet{a15} cannot produce such wide DNS systems, probably requiring 
reconsideration of the binary evolutionary processes. 

The investigation of \citet{k18} involves the amplitudes of the NS natal kicks and the treatment of 
mass and angular momentum transfer. They proposed that a low kick velocity is applied when the envelope 
of the progenitor star is stripped. This can help produce the observed DNS systems with low eccentricities. 
Following the method of \citet{s97} to model the binary orbital evolution with Roche lobe overflow,  
\citet{k18} assumed that the fractions of the mass escaped from the donor and 
ejected from the accretor are respectively 0.2 and $ \geqslant 0.75 $, and 
the rest material (with the fraction of $ \leqslant 0.05 $) is accreted by the accretor. 
In their default model the mass accretion efficiency is also very low, but the channel involving a
double helium star binary was not included. 
For their modelled DNS systems, the probability distribution of the orbital periods is peaked at
$ \sim 1-3 $ days, slightly longer than that ($ \lesssim 1 $ day) predicted in our canonical 
channel. The reason is perhaps that a large fraction of the lost matter from the donor
through a fast wind tend to increase the orbital separations of the progenitor systems (this can lead to more expansion of the donor
and decrease the binding energy of the envelope during CE evolution, broadening the subsequent DNS systems).  
From their orbital parameter distribution (Figure 11), there is a shortage of 
wide systems with orbital period larger than several days with respect to the observed DNSs. 
Meanwhile close systems with orbital period less than $  \sim 1$ day are also deficient. 

\citet{v18} suggested that a bimodal natal kick distribution (with $ \sigma_{\rm k} = 30 \rm\,km\,s^{-1} $
and $ 265 \rm\,km\,s^{-1} $) is preferred over a high-kick component alone. This is in accord with our assumption about
the NS kick velocity distribution. They adopted the same treatments of the NS mass and kick velocities for 
both ultra-stripped SNe \citep[see][]{t15} and ECSs, which can boost 
the formation of DNS binaries with low eccentricities. 
In their fiducial model, the canonical channel is responsible for formation of about 70\% of all DNSs
and the channel with a double core CE phase approximately contributes about 21\%. It can be seen that 
their double helium star channel contributes a significant fraction of DNS population, 
but still much smaller than that in our results. 
Since the probability distribution of the orbital parameters for DNS systems is not clearly shown by \citet{v18}, 
we cannot make further comparisons.

\section{Summary}

We have performed a population synthesis study of the formation of DNS systems in the Galaxy. 
Based on previous investigations on Wolf-Rayet/O binaries, Be/black hole systems and runaway Be stars, 
we argue that mass transfer in the primordial binaries is likely to be highly inefficient. So we adopt the
rotation-dependent mass transfer mode (Mode I) in our calculations. We then
investigate the formation channels of the DNSs and the effect of the SN kicks at the NS formation 
on the distribution of DNS systems. By comparing with observed DNS binaries, we can set important constraints 
on the model parameters. We summarise our main results as follows.
 
1. The second born NSs in most DNS systems have low kick velocities, with the dispersion velocity 
of less than  $ 80 \,\rm km\,s^{-1} $ in a Maxwellian distribution. This also requires that the helium core
mass range for ECS is $ 1.83-2.75 M_{\odot} $ if low kicks for ultra-stripped SNe 
are not taken into account. 

2. The DNS population can be divided into two groups
by the orbital period of $ \sim 1 $~day, the systems with short periods mainly form through the 
canonical channel involving a HMXB phase and the wide binaries experience a 
double helium star binary stage during the progenitor evolution. 

3. Our best model shows that the birthrate and merger rate of Galactic DNS systems 
in the canonical channel are 
$ \sim 4\,\rm Myr^{-1}$ and $ \sim 2\,\rm Myr^{-1}$, and the corresponding rates 
in the double helium star channel are $ \sim 6\,\rm Myr^{-1}$ and $ \sim 2\,\rm Myr^{-1}$,
respectively.

At last we briefly discuss 
the applicability of the $ \Delta M_{\rm ecs} $ that is ascribed to the process of ECS
explosions. Although \citet{a15} extended the upper mass limit to be $ 3.24M_{\odot} $ in their study
on DNS formation, our best range of $ \Delta M_{\rm ecs} = (1.83-2.75 M_{\odot}) $
seems to be still wider than previously expected. Investigations
indicate that, besides ECS, both ultra-stripped SNe and low-mass Fe-core SNe are expected to have considerably lower 
NS kicks than CCS of massive stellar cores \citep{t15,t17,j17,gj18}. This implies that the wide core mass range 
may include the contributions of ultra-stripped SNe and low-mass Fe-core SNe.

\acknowledgements
We thank the referee for constructive comments that helped improve this paper. 
This work was supported by the Natural Science Foundation of China (Nos. 11603010, 11133001, 11333004,
11773015, and 11563003), the Project U1838201 supported by NSFC and CAS, the National Program on 
Key Research and Development Project (Grant No. 2016YFA0400803), and the Natural Science Foundation for 
the Youth of Jiangsu Province (No. BK20160611).


\clearpage

\begin{figure}

\centerline{\includegraphics[scale=0.5]{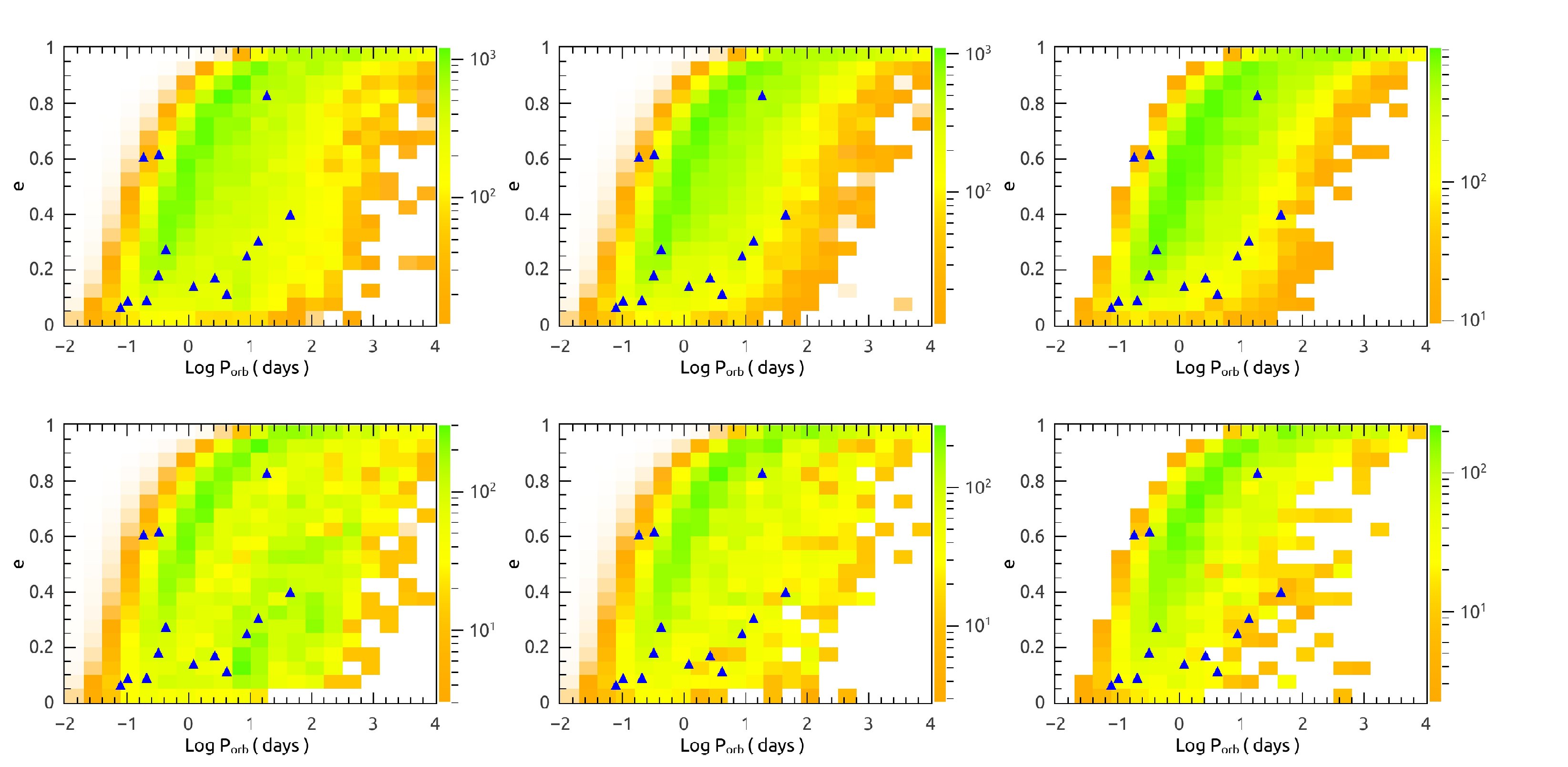}}
\caption{Distributions of Galactic DNS binaries in the orbital period $ P_{\rm orb} $
vs. eccentricity $ e $ plane when taking  $ \Delta M_{\rm ecs} = (1.83-2.25 M_{\odot} )$. 
Each diagram represents the snapshot of the current DNS population with
 a star formation rate of  $ 3 M_{\odot}\,\rm yr^{-1} $ over 10 Gyr. The left, middle
and right panels correspond to $\sigma_{\rm k,ecs} = $ 20, 40 and $80 \rm\, km\, s^{-1} $, 
and the top and bottom panels correspond to $\sigma_{\rm k,ccs} = $ 150 and $300 \rm\, km\, s^{-1} $, respectively.
The blue triangles demonstrate the positions of the observed DNSs.
   \label{figure1}}

\end{figure}

\begin{figure}

\centerline{\includegraphics[scale=0.5]{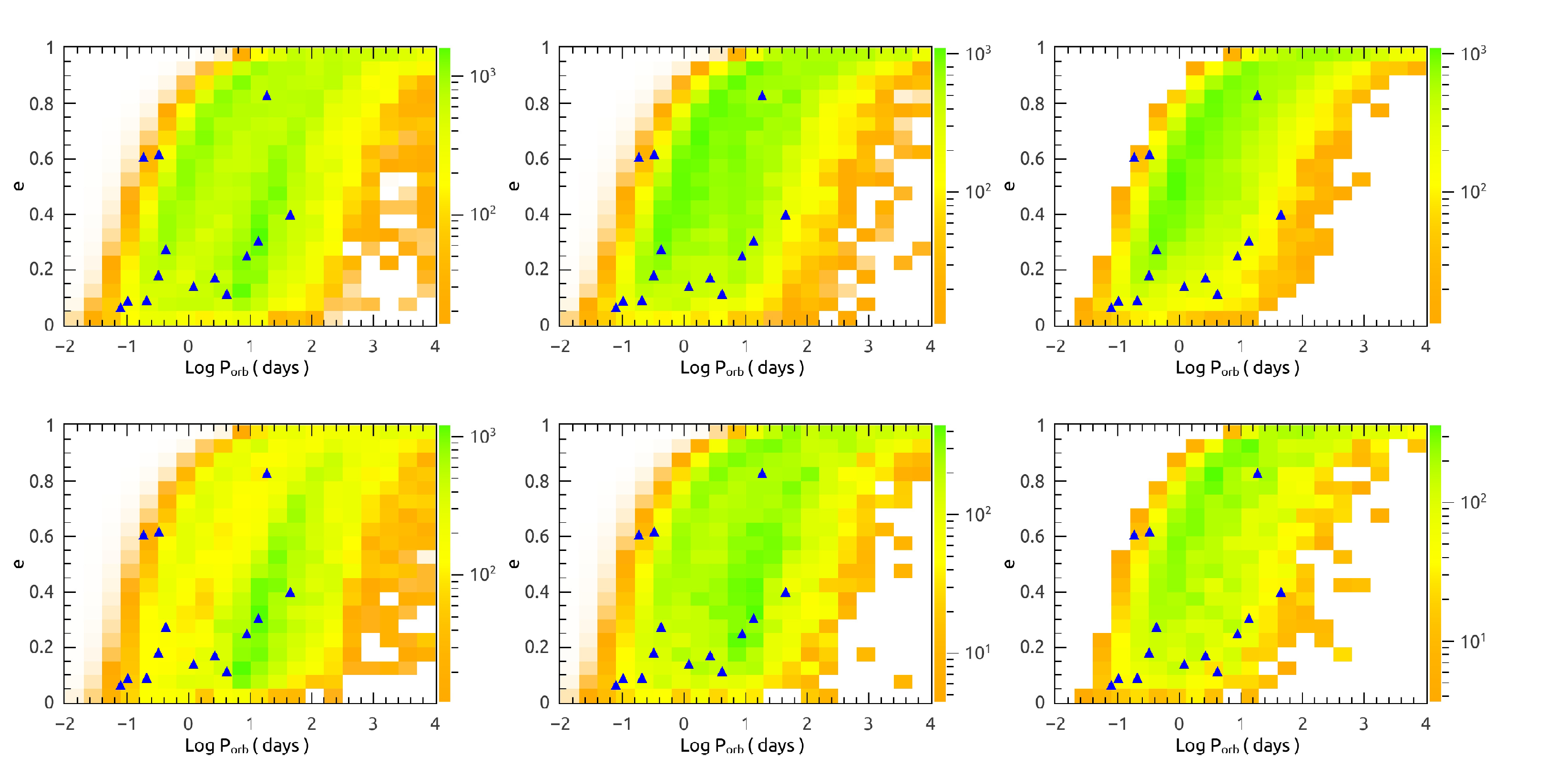}}
\caption{Same as Fig.~1 but with  $ \Delta M_{\rm ecs} = (1.83-2.5 M_{\odot}) $.
 \label{figure2}}

\end{figure}

\begin{figure}

\centerline{\includegraphics[scale=0.5]{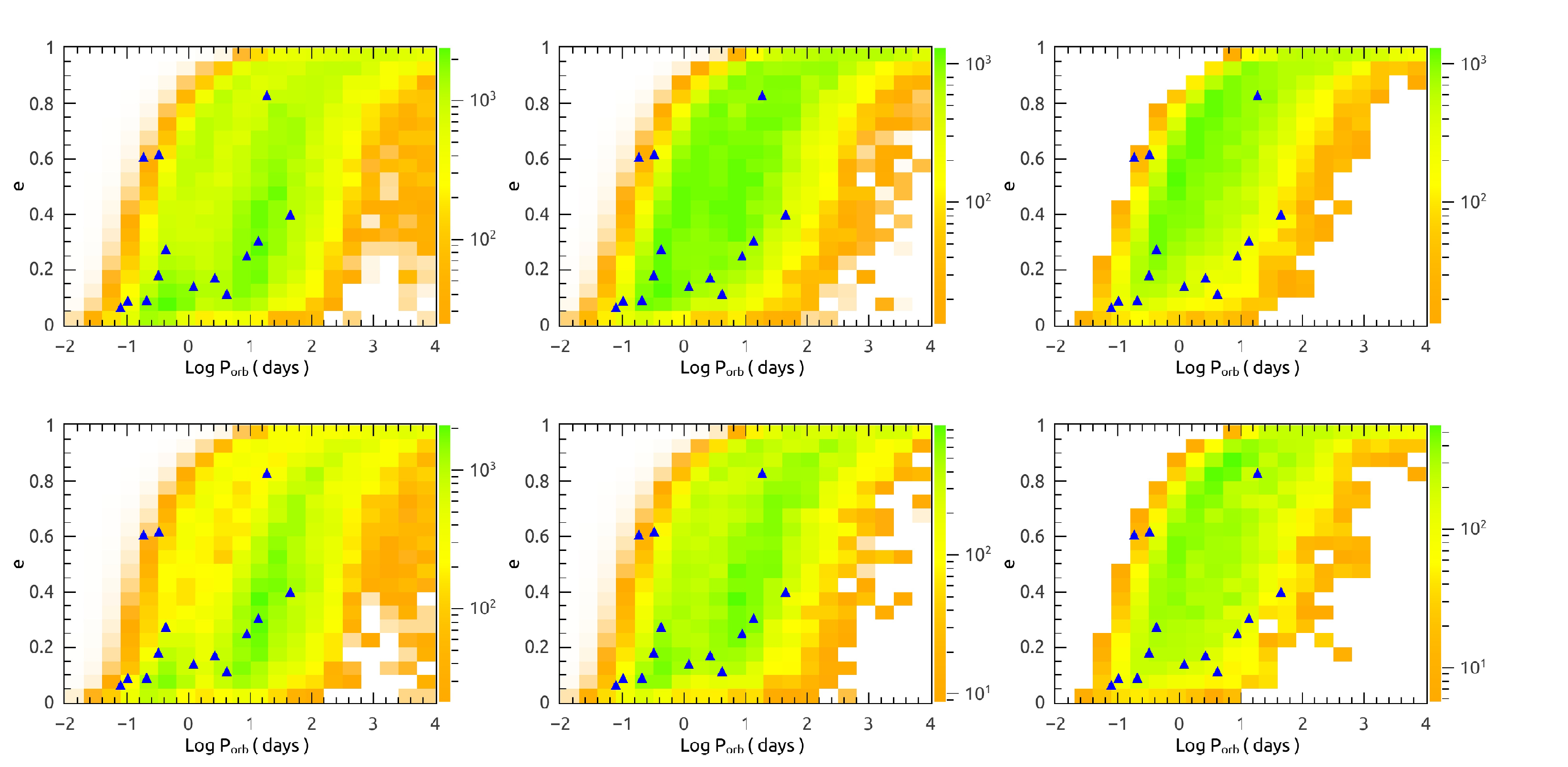}}
\caption{Same as Fig.~1 but with  $ \Delta M_{\rm ecs} = (1.83-2.75 M_{\odot}) $.
\label{figure3}}

\end{figure}

\begin{figure}

\centerline{\includegraphics[scale=0.5]{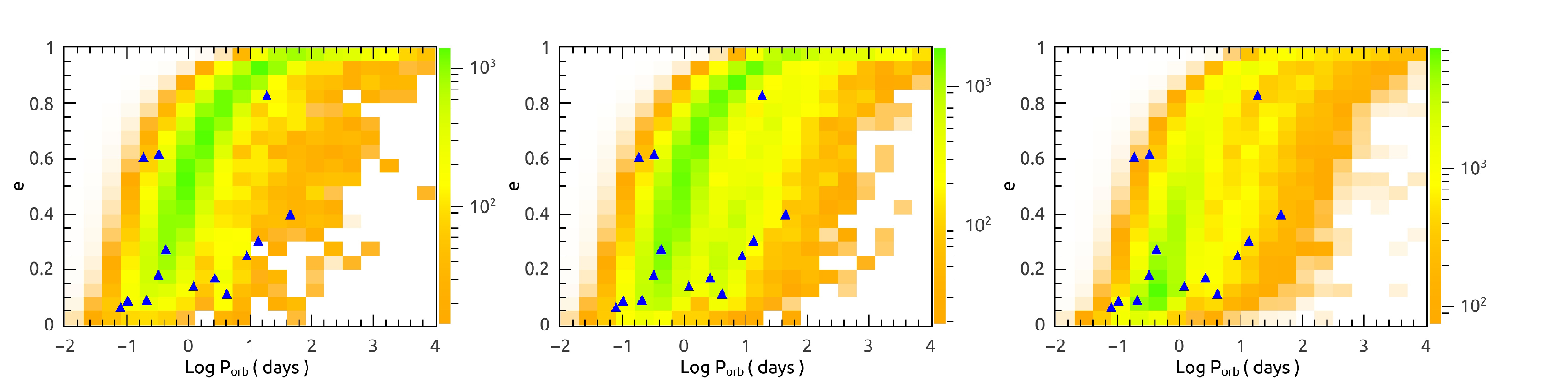}}
\caption{Distributions of Galactic DNS binaries in the  $ P_{\rm orb} - e $ plane under Mode II.
The left, middle and right panels correspond to  $ \Delta M_{\rm ecs} = (1.83-2.25 M_{\odot}) $, 
$ (1.83-2.5 M_{\odot} )$ and $ (1.83-2.75 M_{\odot}) $, respectively. 
The blue triangles show the positions of the observed DNSs.
\label{figure4}}

\end{figure}

\begin{figure}

\centerline{\includegraphics[scale=0.5]{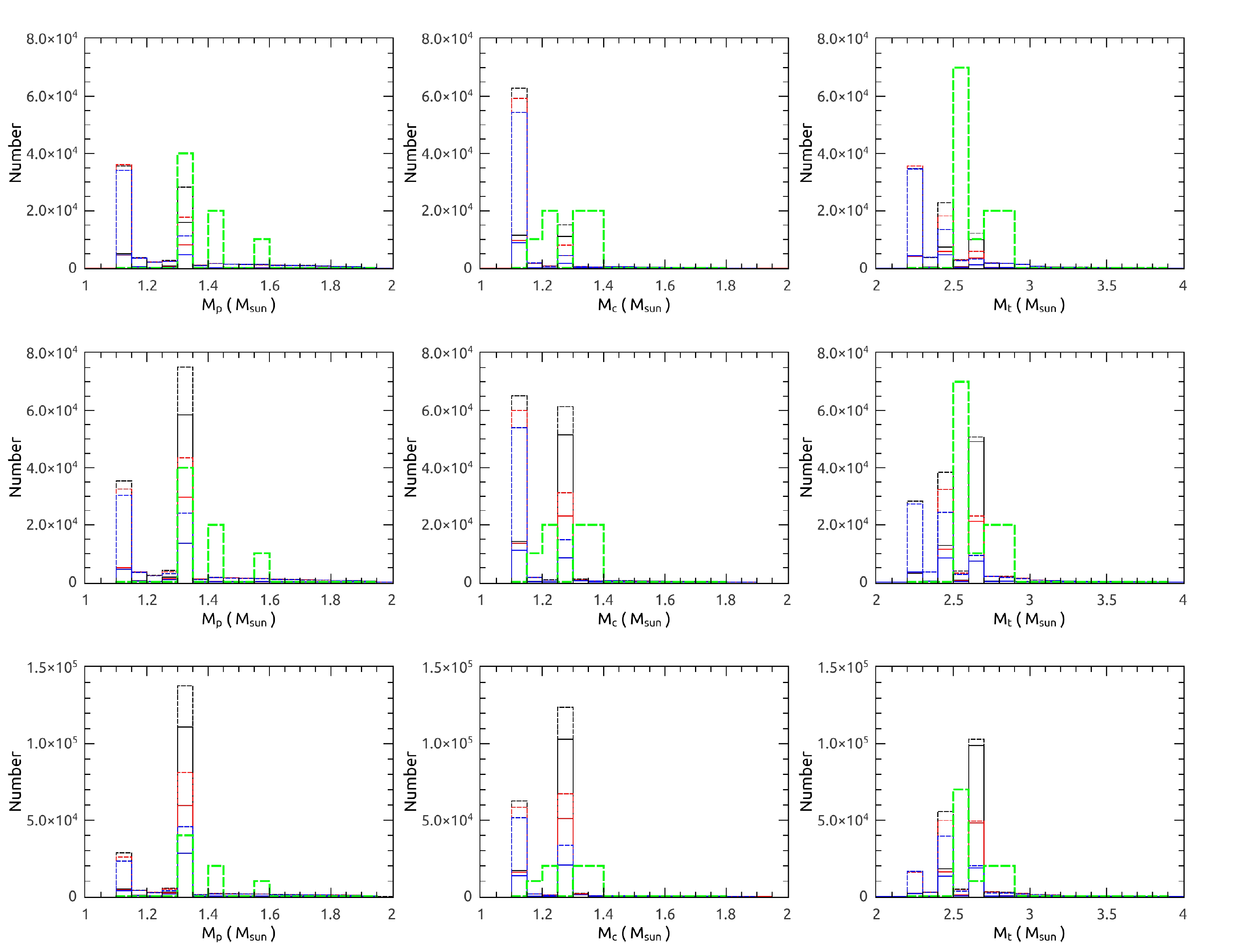}}
\caption{The calculated distributions of pulsar mass $ M_{\rm p} $ (first formed NS, left panels), companion
mass $ M_{\rm c} $ (second formed NS, middle panels) and total mass $ M_{\rm t} $ (right panels) for Galactic DNS binaries. 
The panels from top to bottom correspond to $ \Delta M_{\rm ecs} = (1.83-2.25 M_{\odot} )$, 
$ (1.83-2.5 M_{\odot} )$ and $ (1.83-2.75 M_{\odot} )$, respectively. In each panel, the black, red and blue curves correspond to 
$\sigma_{\rm k,ecs} = $  20, 40 and $80 \rm\, km\, s^{-1} $, and the dashed and solid curves correspond to 
$\sigma_{\rm k,ccs} = $ 150 and $300 \rm\, km\, s^{-1} $ respectively. 
The green curves denote the observed data multiplied by a factor of $ 10^{4} $. There are 7 (13) DNS binaries with  component (total) masses precisely measured (see Table 1). 
\label{figure6}}

\end{figure}

\begin{figure}

\centerline{\includegraphics[scale=0.5]{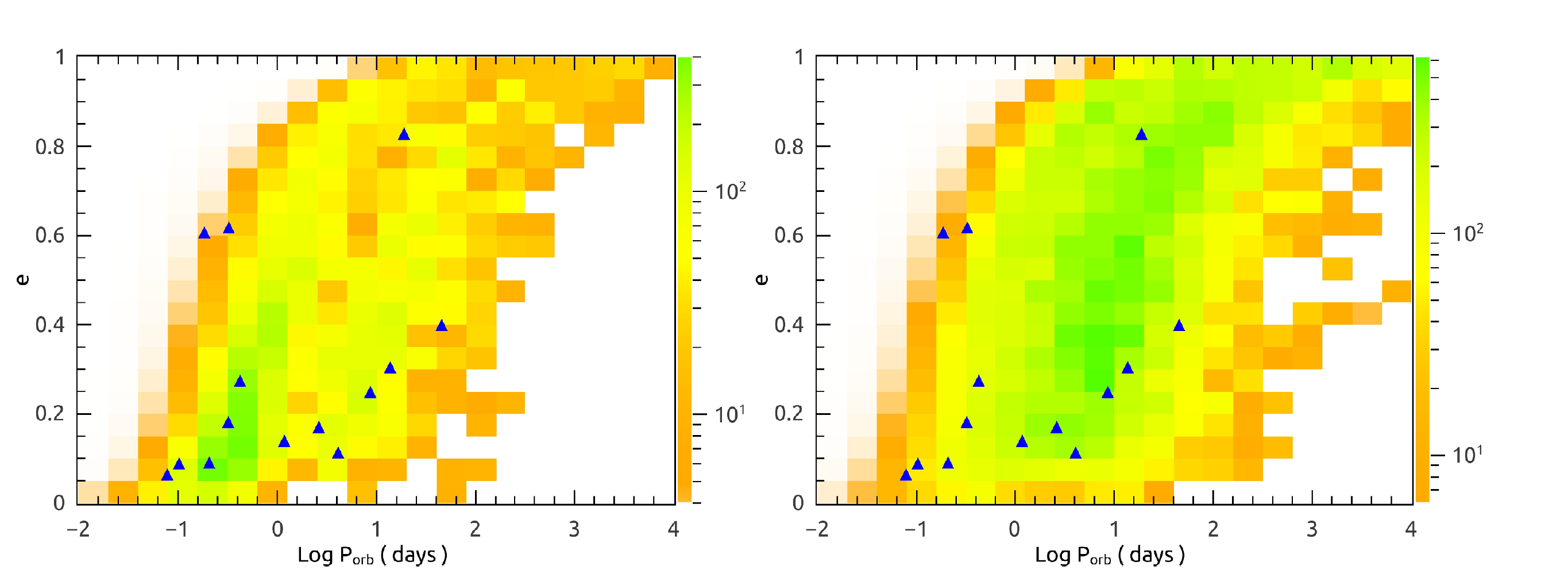}}
\caption{Distributions of Galactic DNS systems formed through the canonical channel (left) and 
the double helium star channel (right). 
Here we adopt the default model with the input parameters of
$\sigma_{\rm k,ecs} = 40 \rm\, km\, s^{-1} $, $\sigma_{\rm k,ccs} = 300 
\rm\, km\, s^{-1} $ and $ \Delta M_{\rm ecs} = (1.83-2.75 M_{\odot}) $.
The observational data are plotted with the blue triangles.
\label{figure6}}

\end{figure}

\begin{figure}

\centerline{\includegraphics[scale=0.5]{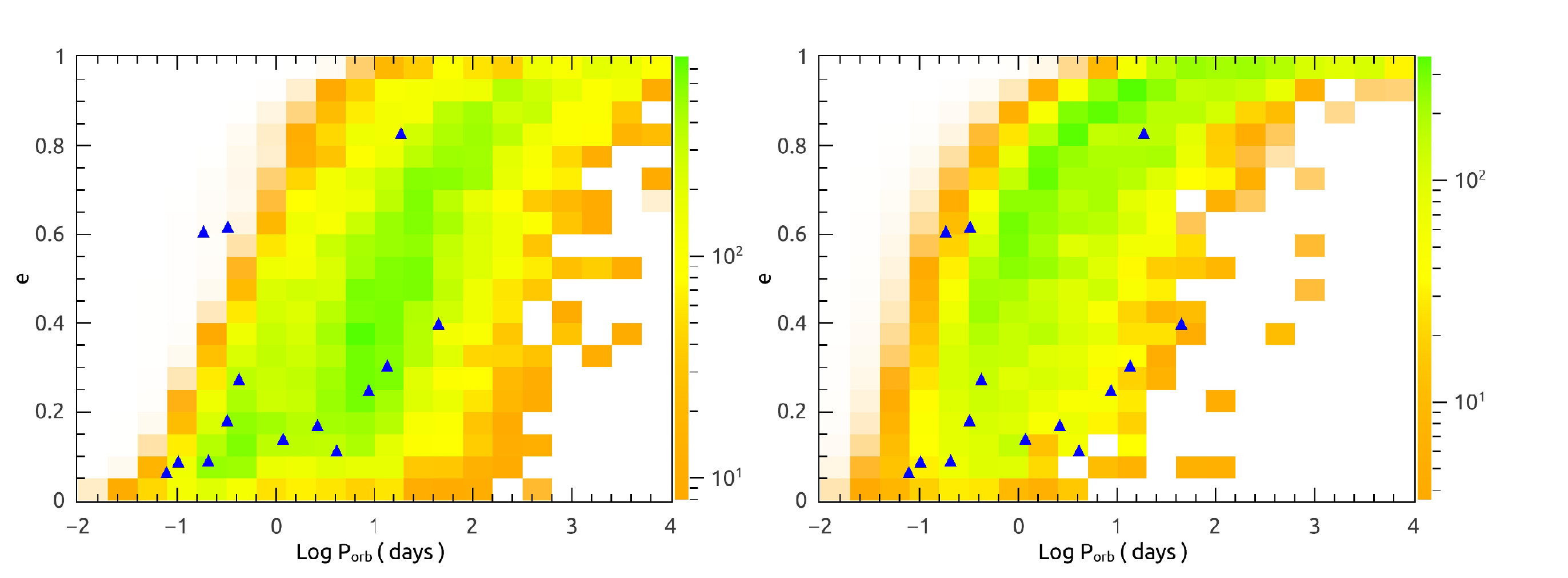}}
\caption{Similar to Figure~6, but with the second born NSs originating from ECS (left) or CCS (right).
\label{figure6}}

\end{figure}

\begin{table}
\begin{center}
\caption{Basic parameters of DNS systems known in our Galaxy, including orbital period $ P_{\rm orb} $,
eccentricity $e$, pulsar mass $ M_{\rm p} $, companion mass $ M_{\rm c} $, total mass $ M_{\rm t} $,
spin period $ P_{\rm spin} $ of recycled pulsars, spin period derivative $ \dot{P}_{\rm spin} $ 
and characteristic age $ \tau_{\rm c} $.
\label{tbl-1}}
\begin{tabular}{lcccccccccclllllll}
\\
\hline
Pulsar name     &  $P_{\rm orb} $ & $e$ & $ M_{\rm p} $ & $ M_{\rm c} $ &  $ M_{\rm t} $ & $ P_{\rm spin} $& $ \dot{P}_{\rm spin} $  &   $ \tau_{\rm c} $ \\
                     &  (days) & & ($M_{\odot}$) &($M_{\odot}$) &($M_{\odot}$) &  (ms)& ($ 10^{-18}\rm s\,s^{-1} $) & (Gyr) \\
\hline
J1946+2052 (1)     &  0.078                 & 0.064  & $< 1.31$& $> 1.18$ & 2.50 &   17.0  &   0.92          & 0.29 \\
J0737-3039A (2)   &  0.102                 & 0.088   & 1.338 & 1.249 & 2.587 &   22.7  &   1.76          & 0.204 \\
J1757-1854 (3)     & 0.184                 &  0.606     & 1.338 & 1.395 & 2.733 & 21.5     & 2.63            & 0.13 \\
J1913+1102 (4)     &  0.206                 & 0.09   & $ <1.84 $ & $ >1.04 $ & 2.88&  27.3    &    0.161      & 2.7 \\
J1756-2251 (5)     &  0.320                 & 0.181   & 1.341 & 1.230 & 2.570 &   28.5    &   1.02        & 0.44 \\
B1913+16 (6)       &  0.323                 & 0.617   & 1.440 & 1.389 & 2.828 &   59.0    &   8.63        & 0.109\\
B1534+12 (7)       &  0.421                 & 0.274   & 1.333 & 1.346 & 2.678&   37.9    &   2.42         & 0.248\\
J1829+2456 (8)     &  1.176                 & 0.139  & $ <1.38 $ & $ >1.22 $& 2.59 &   41.0    &   0.0525    & 12.4\\
J1411+2551 (9)     & 2.616                  & 0.17    & $ <1.62 $ & $ >0.92 $ & 2.54& 62.5    &  0.096       &  10 \\
J0453+1559 (10)     &  4.072                 & 0.113 & 1.559 & 1.174 & 2.734  &   45.8  &   0.186         & 3.9\\
J1518+4904 (11)     &  8.634                 & 0.249   & 1.41& 1.31& 2.72&   40.9  &   0.0272         & 23.9\\
J1753-2240 (12)     &  13.638                 & 0.304 & $ -  $& $ - $ &$ -  $  &   95.1   &   0.97       & 1.55\\
J1811-1736 (13)     &  18.779                 & 0.828  & $ <1.64 $ & $ >0.93 $ & 2.57 &   104.2  &   0.901     & 1.83\\
J1930-1852 (14)     &  45.060                 & 0.399  & $ < 1.32 $ & $ >1.30 $ & 2.59&   185.5   &   18.0     & 0.163\\
\hline
J1906+0746$^{a}$ (15)     &  0.166                  & 0.085 & 1.291 & 1.322& 2.613   &  144.1    &  20300      &  0.0001\\
B2127+11C$^{a}$ (16)     &  0.335                   & 0.681  & 1.358& 1.354 & 2.713 &  30.5     &  4.99        &   0.1 \\
J1755-2550$^{a}$ (17)     &  9.696                   &  0.089 & $ - $ & $ >0.40 $ & $ - $ &  315.2    &  2434     & 0.002 \\
J1807-2500B$^{a}$ (18)    &  9.957                   & 0.747 & 1.366& 1.206 & 2.572 &  4.2  &    0.0823           & 0.81 \\
\hline
\end{tabular}
\end{center}
Note. $ \tau_{\rm c} $ values are calculated with the relation of
 $ P_{\rm spin} / (2\dot{P}_{\rm spin}) $.\\
$^{a}$ These binaries are formed in global clusters (possibly through dynamical encounters) or
not confirmed to be DNSs (maybe contain a white dwarf companion), which are not included in our analyses.\\
Recent references. (1) \citet{s18}. (2) \citet{k06}. (3) \citet{c18}. (4) \citet{l16}. (5) \citet{f14}. (6) \citet{w10}.
(7) \citet{fst14}. (8) \citet{c04}. (9) \citet{m17}. (10) \citet{m15}. (11) \citet{j08}. (12) \citet{k09}.
(13) \citet{c07}. (14) \citet{s15}. (15) \citet{v15}. (16) \citet{j06}. (17) \citet{n18}. (18) \citet{l12}.
\end{table}

\begin{table}
\begin{center}
\caption{Under different assumed models for Galactic DNS population, the obtained  merger rate $ R_{\rm merger} $ 
and  birthrate $ R_{\rm birth} $, the estimated total number, 
and the fraction $ f $ of systems that the second NS formed through ECS among all DNSs. 
The $ \Lambda(P_{\rm orb}, e) $
denotes the likelihood of the observed data for a given model, and the rank of every model is ordered by  
 $ \Lambda(P_{\rm orb}, e)$.
\label{tbl-1}}
\begin{tabular}{ccccccccllll}
\\
\hline
$\Delta M_{\rm ecs}  $  &  $ \sigma_{\rm k,ecs} $ & $ \sigma_{\rm k,ccs} $ 
& $ R_{\rm merger} $   & $ R_{\rm birth} $  & Number  & $ f $  & $ \log(\Lambda(P_{\rm orb}, e)) $ & Rank\\
($ M_{\odot} $) & ($ \rm km\,s^{-1} $)& ($ \rm km\,s^{-1} $) & ($\rm Myr^{-1} $)& ($\rm Myr^{-1} $) 
& ($ \times10^{4} $)   \\
\hline
 1.83$-$2.25  & 20  & 150 & 8 & 15 &  9  & 0.18 &  $-35.7$ & 7\\
 &              & 300 &  3 & 5 & 2   & 0.48 & $-36.1$ & 13\\
 &           40 & 150 &    7 & 13 & 7 &  0.1 & $-36.1$ & 11 \\
 &              & 300 & 2 & 4 & 2 & 0.28 & $-36.1$ & 12\\
 &           80 & 150 &   7  &  12 & 7 & 0.06 & $-36.2$ & 14\\
 &              & 300 &     2  &   3  & 1 & 0.15 & $-37.6$ & 18\\
 1.83$-$2.5   &20 & 150 &    8  &  20   &  13  & 0.45 &  $-35.3$ & 4\\
 &              & 300 &   3  &  9 & 7  & 0.76 &  $-37.0$ & 17\\
 &           40 & 150 &     7  &   16 & 10 &  0.31 &  $-35.6$ & 6\\
 &              & 300 &    3  &  6  & 4 & 0.61 &  $-35.7$ & 8\\
 &           80 & 150 &    7  &  13  & 8 & 0.19 &  $-36.3$ & 15\\
 &              & 300 &     3    &   4   &  2   & 0.41 &  $-36.8$ & 16\\
1.83$-$2.75 &20 & 150 &  10  & 27  & 19  & 0.64 &  $-34.9$ & 3\\
 &             & 300 &  5 & 17  & 12  & 0.85 &  $-35.7$ & 9\\
 &          40 & 150 &   9  &  20  &  13  &  0.5 &  $-34.4$ & 2\\
 &              & 300 &   4  &  10  &  7  & 0.74 &  $-34.1$ & 1 \\
 &           80 & 150 &   8 & 15  & 9  &  0.36 &  $-35.9$ & 10\\
 &              & 300 &  4  & 7  & 4 &  0.58 &  $-35.4$ & 5\\

\hline
\end{tabular}
\end{center}
\end{table}

\end{document}